\documentclass{aa}  

\usepackage{graphicx}
\usepackage{txfonts}
\usepackage{xcolor}
\newcommand{\msun}{{M$_\odot$}}

\newcommand{\Aov}{$\alpha_{\rm ov}$}
\newcommand{\Denv}{log(D$_{\rm env}$/cm$^{2}$s$^{-1}$)}
\begin{document}

\title{Mixing due to internal gravity waves can explain the CNO surface abundances of B-type detached eclipsing binaries and single stars}
\titlerunning{CNO surface abundances and binary evolution}

   \author{Hannah E. Brinkman
          \inst{1}
          \and Andrew Tkachenko
          \inst{1}
          \and Conny Aerts
          \inst{1,2,3}
          }

   \institute{Institute of Astronomy, KU Leuven, Celestijnenlaan 200D, 3001, Leuven, Belgium \and Department of Astrophysics, IMAPP, Radboud University Nijmegen, PO Box 9010, 6500 GL Nijmegen, The Netherlands \and Max Planck Institute for Astronomy, Königstuhl 17, 69117 Heidelberg, Germany
             }
\offprints{hannah.brinkman@kuleuven.be}
   \date{Received 10 June 2025 ; accepted 22 July 2025}
 
  \abstract
   {Observations of double-lined spectroscopic eclipsing binaries are ideal to study stellar evolution. Such stars have tight model-independent constraints on their masses and radii. With the addition of spectroscopically determined effective temperature and surface abundances, they can be used to calibrate and improve stellar evolution models.}
   {The main goal of this work is to determine whether the observed trends of surface nitrogen abundance in single and binary stars can be explained by wave-induced mixing occurring in the stellar envelope.}
   {We use the MESA stellar evolution code to run simulations of single B-type stars with envelope mixing induced by internal gravity waves. We compare the outcome of these models to observations of the surface nitrogen abundance for samples of detached eclipsing binary systems and of single B-type stars. From this comparison we determine 
   the amount of wave-induced mixing required to bring the model predictions in agreement with the observations.}
   {We find nitrogen to be enriched at the surface of theoretical models with wave-induced mixing provided that we use levels above
   \Denv{}=5-6 at the position of the convective core boundary. This corresponds to the highest levels of envelope mixing derived from asteroseismic modelling of single B stars. A prominent observation is that the B-type components of detached eclipsing binaries do not show any nitrogen surface enhancement, which can be explained by their relatively fast rotation enforced by the tidal forces in the systems. The slowly rotating or evolved stars among the sample of single B stars do reveal a nitrogen enhancement. Our findings on the difference in surface nitrogen abundances between single B stars and B-type components of detached binary systems
   can potentially be explained by internal wave-induced mixing profiles based on recent 2-dimensional hydrodynamical simulations of rotating B stars with waves excited at the interface between the convective core and radiative envelope. Such wave-induced mixing decreases with increasing rotation and may act in combination with additional rotational mixing.} 
   {Our findings motivate future asteroseismic studies in large samples of single B stars and pulsating eclipsing binaries with B-type components as optimal laboratories to further test our interpretations in terms of internal wave mixing. }


   \maketitle

\section{Introduction}
Stellar evolution is largely governed by physical processes occurring in stellar internal layers that remain hidden from direct observation due to the high opacity of stellar material. Asteroseismology offers a powerful tool to probe these deep interiors by studying stellar pulsations \citep[see, e.g.][for a recent summary]{Aerts2003, AstroseismologyBook, Montalban2013, Moravveji2015, Deheuvels2016, Mombarg2019, Aerts2019, Viani2020, Noll2021, Pedersen2021, Burssens2023,Aerts2024}.

Recent asteroseismic modelling studies, based on ultra high-quality space-based photometric observations, have revealed discrepancies between observationally inferred interior properties and those predicted by state-of-the-art models of stellar structure and evolution \citep[see, e.g.,][]{Michielsen2019,Michielsen2021,Pedersen2021,Pedersen2022}.
Among the most pressing issues is the amount and functional form of internal mixing, both near the convective core and within the radiative envelope. Asteroseismic investigations of B-type stars born with large convective cores indicate that near-core mixing must be included in models to match observed pulsation frequencies and that a diversity of mixing levels occurs \citep[see, e.g.,][]{Briquet2007, Briquet2012,Degroote2010,Moravveji2015,Moravveji2016,Wu2019,Hendriks2019,Pedersen2021,Burssens2023}. In addition, asteroseismic evidence is mounting for moderate to high levels of mixing in the radiative envelopes of B-type stars, which is needed to explain their observed oscillation properties \citep[see, e.g.,][]{Michielsen2019, Pedersen2021}. While asteroseismology does not yet offer definitive answers regarding the physical origin of these mixing processes, it clearly demonstrates that internal mixing in stars is more efficient than predicted by standard models of stellar structure and evolution without core boundary or envelope mixing. From a multivariate
study, \cite{Aerts2014N14} already concluded from a sample of 68 galactic OB-type stars covering a mass range $[3,40]\,$M$_\odot$ that internal mixing must occur early on in the core-hydrogen burning phase and that it cannot be due to rotational mixing alone.

Detached eclipsing binaries (dEBs) represent an ideal laboratory for studying stellar evolution and the global properties of stellar interior structure \citep[see, e.g.,][]{Guinan2000, Torres2010, Rosu2020, Rosu2022, Rosu2022b}. In general, EBs provide tight constraints on the masses and radii of their components \citep{Torres2010,Serenelli2021}, as well as on their effective temperatures ($T_{\rm eff}$), by breaking the degeneracy between this parameter and surface gravity ($\log\,(g)$) in spectroscopic analyses. The high precision in determining the $T_{\rm eff}$–$\log\,(g)$ parameter pair, in turn, enables accurate inference of the stars’ atmospheric chemical compositions.

Just as asteroseismic studies of single stars do, detailed modelling of eclipsing binaries reveals the need for enhanced near-core mixing resulting in higher convective core masses compared to predictions from standard models of stellar structure and evolution \citep[see, e.g.,][]{OvershootClaret,ClaretTorres2017,Claret2018,Claret2019,Costa2019,Viani2020}. Furthermore, \citet{Andrew2020} argue that substantial convective boundary mixing is required to reconcile observationally determined dynamical model-independent stellar masses with those predicted by evolutionary models. Similar conclusions have been reached by independent studies of eclipsing binaries using isochrone fitting \citep[see, e.g.,][]{Guinan2000, Massey2012, Morrell2014,Johnston2019,Rosu2020, Rosu2022, Rosu2022b}.

Although studies of both single and binary stars agree that additional interior mixing is needed to reconcile observations with theoretical predictions, there are notable areas of disagreement. One of the most striking discrepancies concerns surface nitrogen abundances. While many single stars exhibit nitrogen enrichment at their surfaces \citep[see, e.g.,][]{Hunter2009, Brott2011, Nieva2012, Rivero2012,Bouret2012, Bouret2013, Aerts2014N14}, no similar trend is observed in dEBs. \citet{Pavlovski2018} report a lack of variation in surface nitrogen abundance across a sample of high-mass dEBs, despite the stars spanning a wide range of masses, evolutionary stages along the main sequence, and rotation rates. These findings contrast those of \citet{Nieva2012}, who studied a sample of twenty slowly rotating single B-type stars and found significant variations in surface nitrogen abundance.

The main goal of this work is to determine whether the observed trends of surface nitrogen abundance in single and binary stars can be explained by wave-induced mixing in the envelope rather than by rotationally induced mixing as is commonly assumed.
On the main sequence massive stars burn hydrogen into helium via the CNO cycles. The products of the central hydrogen burning show themselves on the surface of the stars, though the time scale on which this happens depends on the internal mixing processes in the envelope. \cite{Brinkman2024} investigated the impact of mixing caused by internal gravity waves (hereafter IGWs) on the evolution of a 20\,\msun{} single star. Here, we apply a similar approach to the components of the binary systems presented in \citet{Andrew2020}, by simulating single star tracks and adding envelope mixing based on IGWs to investigate its impact on the surface abundances of C, N, and O. Our focus on IGW mixing is inspired by the results from \cite{Pedersen2021, Pedersen2022}, who investigated the amount of envelope mixing needed to match asteroseismic observations of B-type stars. They found a wide range for the efficiency of envelope mixing, $D_{\rm env}$, with values between roughly 10 and $10^6\,$cm$^2$\,s$^{-1}$. Moreover, numerical hydrodynamical simulations of IGW 
by \citet{Rogers2013} offered the appropriate functional form of the mixing
they induce following the initial description based on particle diffusion by \citet{Rogers2017}, and its generalisations for a variety of masses \citep{Varghese2023} and internal rotation rates \citep{Varghese2024}. 
In particular \citet{Varghese2024} highlighted the strong decrease in efficiency due to IGW mixing as the star's convective core shrinks while rotating faster as they evolve along the main sequence. This is due to the decrease in stochastic excitation of the IGW throughout the evolution. 
However, when relying on these results from simulations, we must keep in mind that angular momentum transport is still poorly understood for B-type stars during their main sequence \citep{Aerts2019}. Asteroseismology points out that the shrinking core of stars do not necessarily get faster rotation as they approach the end of the main sequence \citep{Aerts2025}. 

 To gain more insight in the expected behaviour of the surface abundances of C, N, and O, we also look at the surface abundances of several sets of observed massive stars from the literature. Massive main-sequence stars with a mass below 20\,\msun{} are ideal for this purpose since they do not migrate far from their birth environment due to their short lifetimes. For masses below 20\,\msun{}, the stellar winds are weak enough that they do not strongly affect the photosphere \citep{Przybilla2013}. Hence any change in the surface abundances is interpreted as the result of efficient internal mixing, whether it is induced by IGWs or by stellar rotation \citep[see, e.g.,][]{MandM2000, MandM2000b, Heger2000b, Brott2011} or a combination thereof, rather than exposure of processed material by stripping of the envelope as occurs in stars that have undergone binary interactions \citep[see, e.g.,][]{Podsiadlowski1992, Langer2003, Tauris2015,Ma2025}.
 
The structure of the paper is as follows: in Section \ref{Method} we describe the input of the stellar evolution models used for this work. In Section \ref{discrepancy} we compare the stellar evolution tracks of our models with observations for the binary systems. In section \ref{Surface} we first describe the theory and the theoretical boundaries for the CNO equilibrium and then compare the results for the binaries to those of several samples of observed massive stars in the literature. We describe the uncertainties for the surface abundance coming from the stellar evolution models and compare them to those of the observations. In Section \ref{Observations} we discuss the differences between the models and the observations and a way forward to resolve them. We finish with our conclusions in Section \ref{Conclusions}.

\section{Stellar models with IGW envelope mixing}\label{Method}
We have used the MESA (Modules for Experiments in Astrophysics) stellar evolution code version 22.11.1 \citep{MESA1,MESA2,MESA3,MESA4, MESA5, MESA6} for the simulations presented in this paper. We followed the same model setup as in \cite{Brinkman2024}. We repeat the main points here but refer to \cite{Brinkman2024} for more details.

The nuclear reaction rates were taken from the JINA Reaclib \citep{CyburtJINA2010}. The initial mass of the models is in the range 5-20\,\msun{}. We used an initial metallicity of Z=0.014 (current Solar) combined with the chemical mixture as given by \cite{Przybilla2013}. The initial hydrogen content is then defined as X$_{ini}$=1-Y$_{ini}$-Z$_{ini}$, where Y$_{ini}$=0.2465+2.1$\times$Z$_{ini}$. The value of 0.2465 refers to the primordial helium abundance as determined by \cite{Aver2013}. The value of 2.1 is chosen such that the mass fractions of the chemical mixture adopted from \cite{Przybilla2013} are reproduced \citep[see also][]{Michielsen2021}. The nuclear network contains all the relevant isotopes for the main burning phases (H, He, C, Ne, O, and Si), allowing to follow the evolution of the stars in detail up to core collapse, though here the focus is on the main sequence, for more details see \cite{Brinkman2024}.

\begin{figure}
    \centering
    \includegraphics[width=\linewidth]{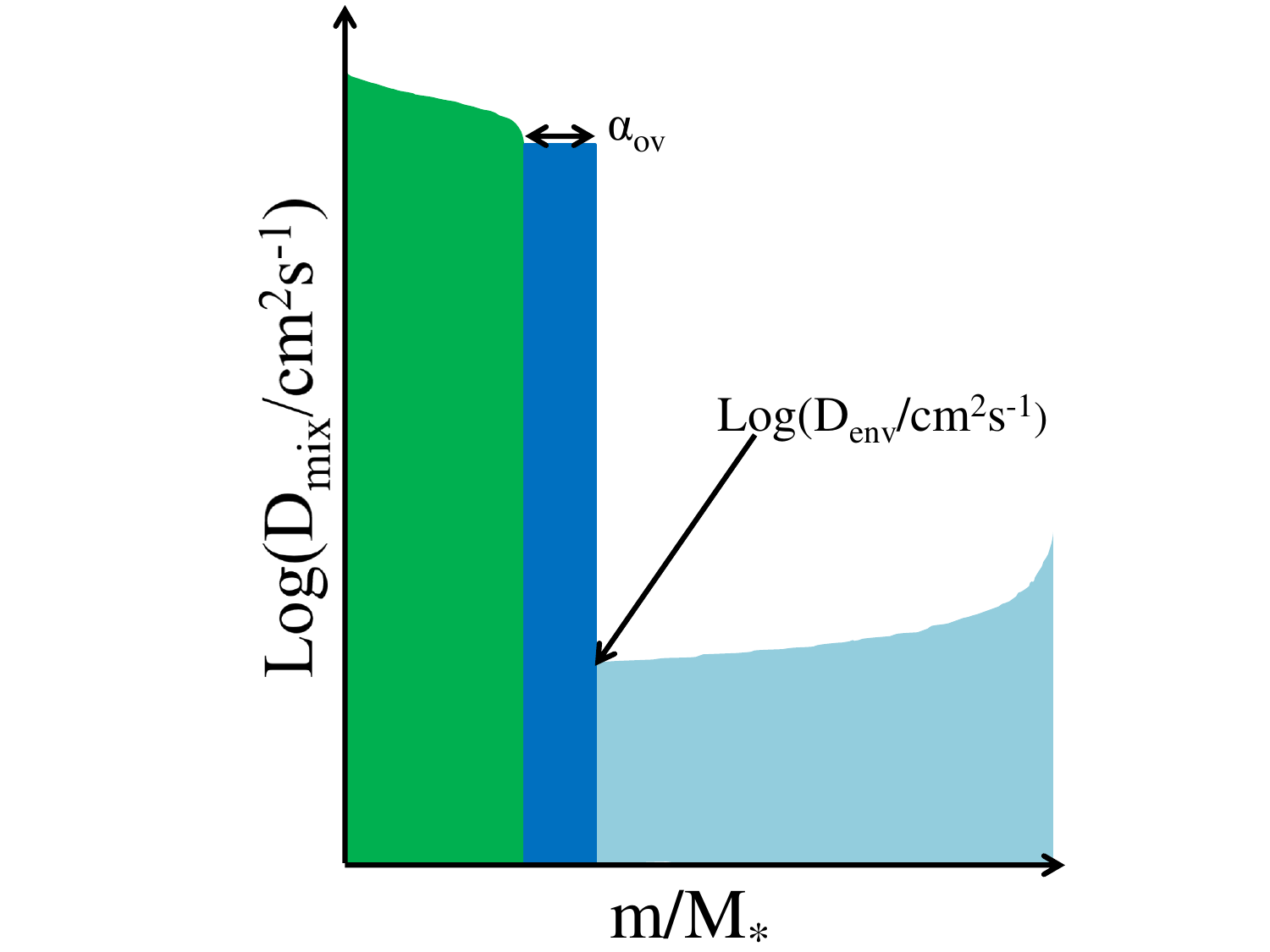}
    \caption{Shape of the overall internal mixing profile in our stellar models. The green zone on the left is the fully mixed convective core, the dark blue zone indicates the convective overshoot zone, and the light blue area is the radiative envelope where mixing takes place through IGWs. The size of the overshoot zone is determined by $\alpha_{ov}$, and the starting height of the envelope mixing is determined by \Denv{}. This figure is inspired by Figure 3 of \cite{Pedersen2021}.}
    \label{Mixing}
\end{figure}

We made use of the Ledoux criterion to establish the location of the convective boundaries. Convection itself was treated according to the MLT++ prescription \citep{MESA2}. This prescription introduced a free parameter, $\alpha_{mlt}$. Here we set $\alpha_{mlt}$ to 2 for all phases of the evolution as a suitable value for main-sequence B stars.
We adopt the CBM scheme as implemented by \cite{Michielsen2019}, and we make use of the so-called `step-overshoot' part of this scheme (also referred to as convective penetration). We kept the CBM, parametrised by \Aov{}, fixed to a value of 0.2, which is commonly used for massive stars. To determine the location of the convective boundary we enhanced the mesh locally around the hydrogen burning core following asteroseismic modelling of B stars \citep{Pedersen2021,Burssens2023}. Extra mixing in the envelope is needed for the stellar evolution models to match the asteroseismic observations. In this work we focused on the profile due to IGW mixing as implemented by \cite{Michielsen2021}. 
This profile is inspired by the simulations in \citet{Rogers2017}, who found a radial profile proportional to a function between $1/\rho$ and $1/\sqrt{\rho}$. Concretely, the functional form we used is given by  D$_{IGW}$(r)=D$_{env}$($\rho(r_{CBM})$/$\rho(r)$), where D$_{env}$ is a free parameter  determining the strength of the mixing at the connection point between the extended CBM region and the envelope, $\rho(r)$ is the local density with
$\rho(r_{CBM})$ being the density at the border between the CBM region and the envelope. We represent the overall shape of the mixing profile in the models in Fig.\,\ref{Mixing}. The diffusion coefficient, \Denv{}, was varied from 0 to a maximum value to achieve changes in the surface abundances. This maximum value is based on the impact on the stellar evolution and the range found by \cite{Pedersen2022}. For initial stellar masses up to 10\,\msun{} the maximal envelope mixing is set to \Denv{}=5, while for initial masses above 10\,\msun{} 
the maximum is set to \Denv{}=6. This latter upper limit is taken because higher values of the envelope mixing lead to a chemically homogeneous evolution, which is not observed for the stars presented here.

Aside from stellar models with IGW envelope mixing, we
also simulated a few model tracks with rotationally induced mixing adopting the implementation in MESA. In doing so we fixed the free parameters set to the standard values in \cite{Heger2000}. The models presented below either have envelope mixing induced by IGWs or by rotation, but not a combination of the two.
We have evolved the stars to the end of the main sequence, that is, before mass loss starts to play a significant role in the changes of the surface abundances.

\section{The binary sample}\label{discrepancy}
\begin{figure}
    \centering
    \includegraphics[width=\linewidth]{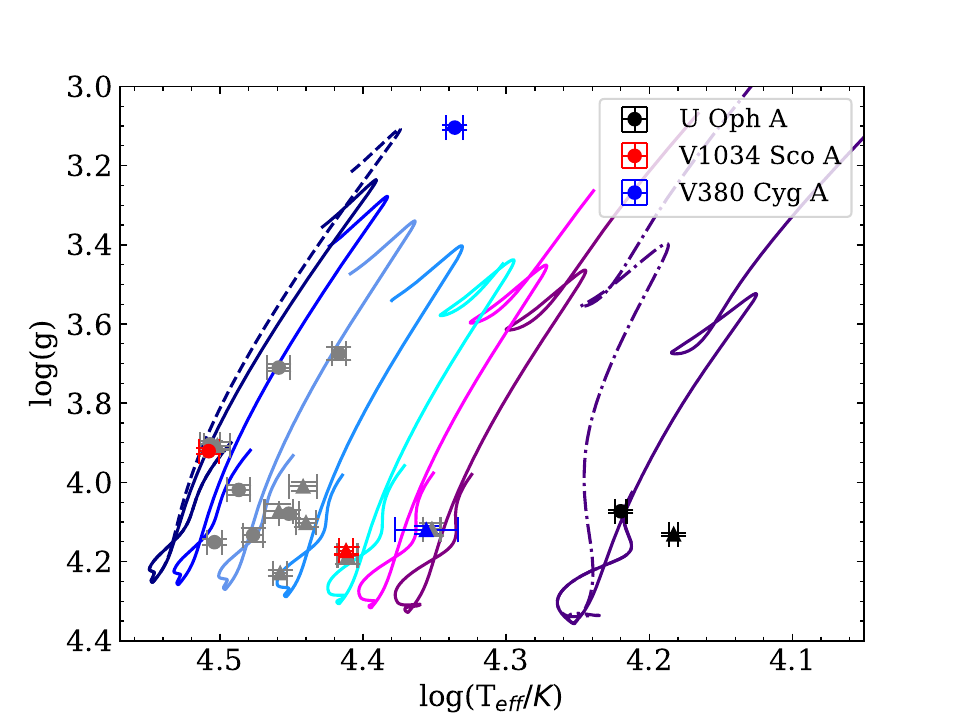}
    \caption{Kiel diagram for the components of the binary systems from \cite{Andrew2020} along with the stellar evolution tracks of models with masses from 5-20\,\msun{}. For 5 and 20\,\msun{} the tracks with the maximum amount of IGW envelope mixing, \Denv{}=5 and 6, are shown with the dashed-dotted and dashed lines, respectively. None of the models in this figure include the effects of rotational mixing. The binaries discussed in Section \ref{discrepancy} are indicated by the black symbols (U Oph), red symbols (V1034 Sco), and blue symbols (V380 Cyg). The stellar parameters of the binaries are reported in Table \ref{BinaryData}.}
    \label{KielBinary}
\end{figure}

In this study, we focus on the sample of dEBs presented in \citet{Andrew2020}. The properties of the sample are summarized in Table\,\ref{BinaryData}. We compare the stellar evolution tracks of single stars with the observed effective temperature and determined surface gravity of the components of the binary systems in Figure~\ref{KielBinary}. For 5 and 20\,\msun{} models, we feature tracks corresponding to no mixing (solid lines) and the maximum adopted amount of envelope mixing due to IGWs for a given mass (dash-dotted and dashed lines). Because the binaries are all in a detached state and have not interacted, we assume that we can treat the components as single stars. Furthermore, we assume that the surface gravity is both precise and accurate for all stars in the sample, as is also the case for their observationally inferred masses.

 The binary sample was discussed by \citet{Andrew2020} in the context of the mass discrepancy problem. This refers to the difference between stellar masses determined in a model-independent way from binary dynamics and those obtained by fitting evolutionary tracks to stellar positions in the Hertzsprung–Russell or Kiel diagrams. Although some studies claim that the mass discrepancy increases with stellar mass \citep[e.g.,][]{ClaretOvershoot2007,OvershootClaret,ClaretTorres2017,Viani2020}, \citet{Andrew2020} concluded that the discrepancy is more pronounced in stars that are more evolved along the main sequence. The authors demonstrated that, at the level of the sample as a whole, the discrepancy can be resolved by including near-core mixing in the form of convective overshooting in the models. This suggests that standard models of stellar structure and evolution may underestimate the extent and mass of convective cores in such stars.
 
A qualitative comparison between the grid of MESA models presented in Section~\ref{Method} and the stellar positions in the Kiel diagram shown in Figure~\ref{KielBinary} allows us to explore whether IGW envelope mixing could serve as an alternative explanation for the discrepancies between models and observations. While a detailed investigation of the mass discrepancy problem lies beyond the scope of this paper, we summarise the conclusions of our qualitative experiment using three representative systems as case studies.

First, we find that the mass discrepancy is either absent or significantly less pronounced in stars located near the zero-age main-sequence (ZAMS). We also conclude that, even if a mass discrepancy were present at the early stages of main-sequence evolution, adjusting the amount of internal mixing -— whether near-core or in the envelope -— would not resolve it. This is best illustrated by the components of the U Oph system (black symbols in Fig.\,\ref{KielBinary}), which are among the lowest-mass and least evolved binary components in the sample. In these stars, any internal mixing, whether driven by IGWs or by rotationally induced processes, has not yet had time to produce a significant effect.

Second, the effect of increasing envelope mixing becomes progressively smaller with increasing stellar mass. As a result, explaining the observed discrepancies through additional envelope mixing becomes increasingly difficult for higher-mass stars. This trend is best illustrated by the V1034 Sco system (red symbols in Fig.\,\ref{KielBinary}), which lies at the upper end of the mass spectrum in our sample. To reconcile the observed and theoretically predicted masses for this system, one would need to assume higher levels of envelope mixing, that is \Denv{} > 6, 
approaching the regime of chemically homogeneous evolution.

Third, we find that the mass discrepancy is largest and least resolved through the inclusion of IGW envelope mixing for the most evolved stars in the sample. In this context, the primary component of the V380 Cyg system (blue symbols in Fig.\,\ref{KielBinary}) serves as an ideal test case, as the star is located near the terminal-age main sequence (TAMS) in the Kiel diagram. As with the previously discussed most massive binary in the sample, reconciling the observed and theoretical masses for V380 Cyg would require an extreme level of envelope mixing. Yet even then, the mass discrepancy cannot be fully resolved.

Based on the qualitative analysis and the conclusions drawn from it, we pose the following question: are such high levels of IGW envelope mixing physically justifiable, and is there an observationally oriented metric beyond the mass discrepancy that could help us to quantify the amount of mixing? We address this question in the following sections.
\begin{table*}
    \centering
        \caption{Selected data for the binary systems from \cite{Andrew2020} supplemented with the surface abundance values from $^{(1)}$\cite{Pavlovski2018}, $^{(2)}$\cite{Pavlovski2023} and $^{(3)}$\cite{Andrew2014}. The error bars are given in the parentheses in terms of the last digit(s). For each object, the first line corresponds to the primary component and the second line corresponds to the secondary component.}
    \begin{tabular}{ccccccccccc}
         Object     &  M        &log T$_{eff}$   &log g   &v sin i    &v$_{eq}$/v$_{crit}$    &log$\epsilon$(C)   &log$\epsilon$(N)   &log$\epsilon$(O)  &[N/C]   &[N/O]\\
                    &  (\msun)  &(dex)           &(dex)         &(km s$^{-1}$)             &($\%$) &               &\\
        \hline                      
        V578 Mon$^{(1)}$    & 14.54(8)  &4.477(7)       &4.133(18)     &117(4)        &21.0(9)          &8.18(7)   &7.69(12)  &8.74(10)    &-0.49(14)   &-1.05(16)\\
                          & 10.29(6)  &4.411(7)       &4.185(21)     &94(2)         &17.9(7)          &8.21(11)  &7.72(9)   &8.76(11)    &-0.49(14)   &-1.04(14)\\
        V453 Cyg$^{(1)}$    & 13.90(23) &4.459(7)       &3.710(9)      &107.2(2.8)    &23.6(1.0)        &8.28(4)   &7.73(9)   &8.74(9)     &-0.55(10)   &-1.01(13)\\
                          & 11.06(18) &4.442(10)      &4.010(12)     &98.3(3.7)     &19.3(1.1)        &8.24(6)   &7.76(8)   &8.74(8)     &-0.48(10)   &-0.98(13)\\
        V478 Cyg$^{(1)}$    & 15.40(38) &4.507(7)       &3.904(9)      &129.1(3.6)    &25.4(1.3)        &8.24(9)   &7.69(11)  &8.65(8)     &-0.55(15)   &-1.03(17)\\
                          & 15.02(35) &4.502(9)       &3.907(10)     &127.0(3.5)    &25.1(1.3)        &8.28(8)   &7.68(9)   &8.72(11)    &-0.60(12)   &-0.99(12)\\
        AH Cep$^{(1)}$      & 16.14(26) &4.487(8)       &4.019(12)     &172.1(2.1)    &32.6(9)          &8.37(8)   &7.64(7)   &8.66(12)    &-0.73(11)   &-1.02(14)\\
                          & 13.69(21) &4.459(10)      &4.073(18)     &160.6(2.3)    &30.9(9)          &8.27(5)   &7.67(6)   &8.69(12)    &-0.60(8)    &-1.02(13)\\
        V346 Cen$^{(2)}$    & 11.78(13) &4.417(5)       &3.675(17)     &165.2(2.8)    &39.0(1.3)        &8.13(5)   &7.68(3)   &8.70(3)     &-0.45(3)    &-1.02(2)\\
                          & 8.40(10)  &4.352(6)       &4.118(16)     &89.1(2.3)     &17.7(7)          &8.33(5)   &7.72(4)   &8.80(1)     &-0.61(3)    &-1.08(2)\\
        V573 Car$^{(2)}$    & 15.14(39) &4.504(5)       &4.151(7)      &184.6(2.7)    &31.4(1.1)        &8.19(8)   &7.61(3)   &8.65(4)     &-0.57(4)    &-1.03(3)\\
                          & 12.38(20) &4.458(5)       &4.229(7)      &155.4(3.1)    &26.5(9)          &8.27(7)   &7.72(2)   &8.67(3)     &-0.55(3)    &-0.94(2)\\
        V1034 Sco$^{(2)}$   & 17.07(12) &4.508(7)       &3.921(8)      &169.8(2.6)    &31.9(8)          &8.39(6)   &7.71(6)   &8.76(2)     &-0.68(4)    &-1.05(3)\\
                          & 9.60(5)   &4.412(5)       &4.173(9)      &94.5(3.3)     &17.8(7)          &8.29(4)   &7.72(1)   &8.74(2)     &-0.57(2)    &-1.02(1)\\
        V380 Cyg$^{(2)}$    & 11.43(19) &4.336(6)       &3.104(6)      &98(2)         &32.6(1.1)        &8.20(2)   &7.55(5)   &8.64(7)     &-0.65(5)    &-1.09(9)\\
                          & 7.00(14)  &4.356(22)      &4.120(11)     &38(2)         &8.0(6)           &8.19(4)   &7.53(8)   &8.55(12)    &-0.66(9)    &-1.02(15)\\
        CW Cep$^{(3)}$      & 13.00(7)  &4.452(7)       &4.079(10)     &105.2(2.1)    &19.3(5)          &8.30(7)   &7.79(8)   &8.71(7)     &-0.51(11)   &-0.92(11)\\
                          & 11.94(7)  &4.440(7)       &4.102(10)     &96.2(1.9)     &17.8(6)          &8.24(7)   &7.70(8)   &8.70(6)     &-0.54(11)   &-1.00(10)\\
        U Oph             & 5.09(5)   &4.220(4)       &4.073(4)      &110(6)        &25.4(1.6)        & -        &-         &-           &-           &-\\
                          & 4.58(5)   &4.183(3)       &4.131(4)      &108(6)        &24.7(1.2)        & -        & -        &-           &-           &-\\
     \hline
    \end{tabular}
    \label{BinaryData}
\end{table*}

\section{IGW versus rotational mixing and surface abundances}\label{Surface}

As a recap from last section, a large amount of envelope mixing would be required if we wanted to match, or at least get close to, the observationally determined positions of the binary components in the Kiel diagram (see Fig.\,\ref{KielBinary}). In this section, we describe what the impact of such amounts of envelope mixing due to IGWs would be on the surface abundances. We compare these theoretical predictions with the observationally determined surface abundances of the binary stars, as well as with the abundances of several samples of single B-type stars collected from the literature. By comparing the different observables, we try to determine whether the mixing we have considered earlier is realistic or not. We focus on the elements of the CNO cycles, most specifically nitrogen.

\subsection{The theoretical CNO equilibrium}\label{Equilibrium}
Of all the reactions making up the CNO-cycles, the $^{14}$N(p,$\gamma$)$^{15}$O is the slowest reaction and thus the amount of $^{14}$N will increase over the duration of the main sequence for the mass regime considered here. In the stellar core, while the amount of nitrogen increases, carbon and oxygen will decrease to an equilibrium value, though the overall amount of C+N+O will remain constant. If enough mixing beyond the convective core is present, $^{14}$N is transported up to the surface, increasing the surface abundance of nitrogen. Carbon and oxygen will decrease slightly, though mostly at the very end of the main sequence evolution. Changes in the surface abundances can also occur due to stripping of the envelope, either through stellar winds or binary interactions. However, the stars in our binary sample did not interact yet and their masses are too low for strong stellar winds to affect them at this stage. Hence all changes on the surface must come from internal mixing.

\begin{figure}
    \centering
    \includegraphics[width=\linewidth]{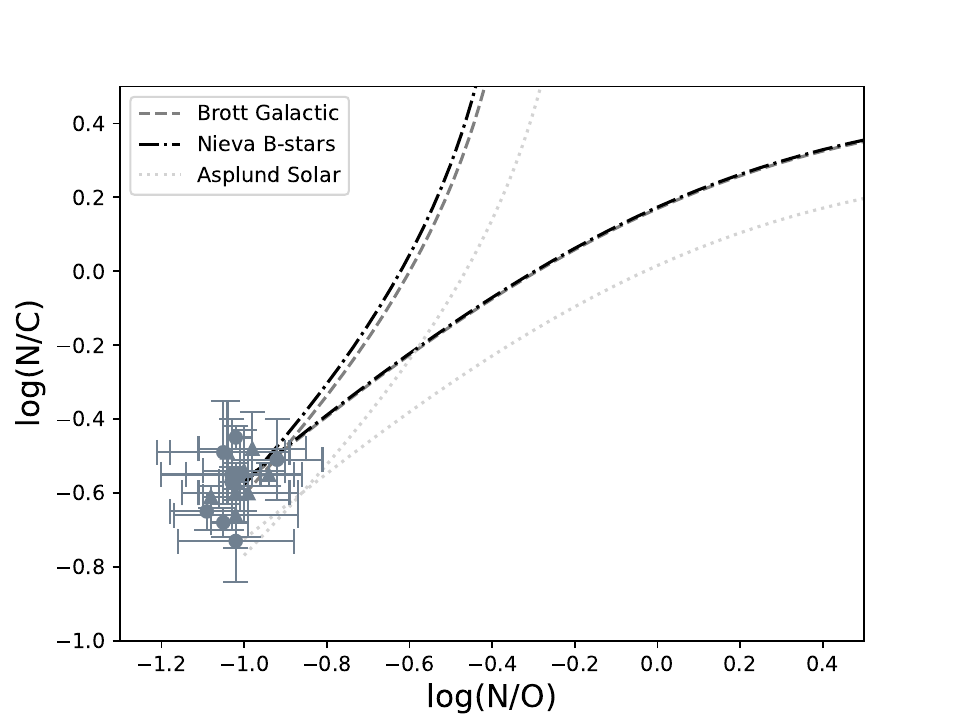}
    \caption{Theoretical boundaries of N/C and N/O based on changes in the CNO-ratio following the description given by \cite{Martins2015CNO} for the elemental mixtures adopted by \cite{Brott2011} (grey dashed line), \cite{Nieva2012} (black dashed-dotted line), and \cite{Asplund2009} (light grey dotted line).  The position of the binary star from \cite{Andrew2020} are shown with their respective error bars.}
    \label{CNOLimits}
\end{figure}

Because of the catalytic nature of the CNO-cycles, theoretical boundaries for the equilibrium fractions of N/O and N/C can be defined \citep[see, e.g.,][]{Maeder2014, Martins2015CNO}. Figure \ref{CNOLimits} shows these theoretical CNO equilibrium boundaries. For the elemental mixtures for galactic sources adopted by \cite{Brott2011} and for B stars in the solar neighbourhood from \cite{Nieva2012}, there are no large differences in the equilibrium boundaries. However, when using the mixture following \cite{Asplund2009} based on the current Sun, the boundaries have an offset compared to the other two mixtures. It is important to note this shift between the mixtures, since it shows that the CNO abundances and thus their ratios are different for different assumptions for the chemical mixture, which depends on the location of the star in the galactic disk and its direct surroundings at birth.

\subsection{Sample stars in the CNO equilibrium diagram}
When projecting the binary stars onto the CNO equilibrium (see Fig.\,\ref{CNOLimits}), all the stars are clustered at the crossing point of the equilibrium lines, at log(N/O)=-1.0 and log(N/C)=-0.5. This means that the binary components, even those that have already evolved away from the ZAMS, do not show any surface enhancement of $^{14}$N. This is in agreement with the relatively modest levels of envelope mixing found from asteroseismic modelling of most of the single B stars \citep{Pedersen2021}. Indeed, their dominant internal mixing occurs in the transition layer between the convective core and radiative envelope as highlighted in the cartoon in Fig.\,\ref{Mixing} and summarized in \citet[][Table\,1]{Aerts2021}. 

\begin{figure}
    \centering
    \includegraphics[width=\linewidth]{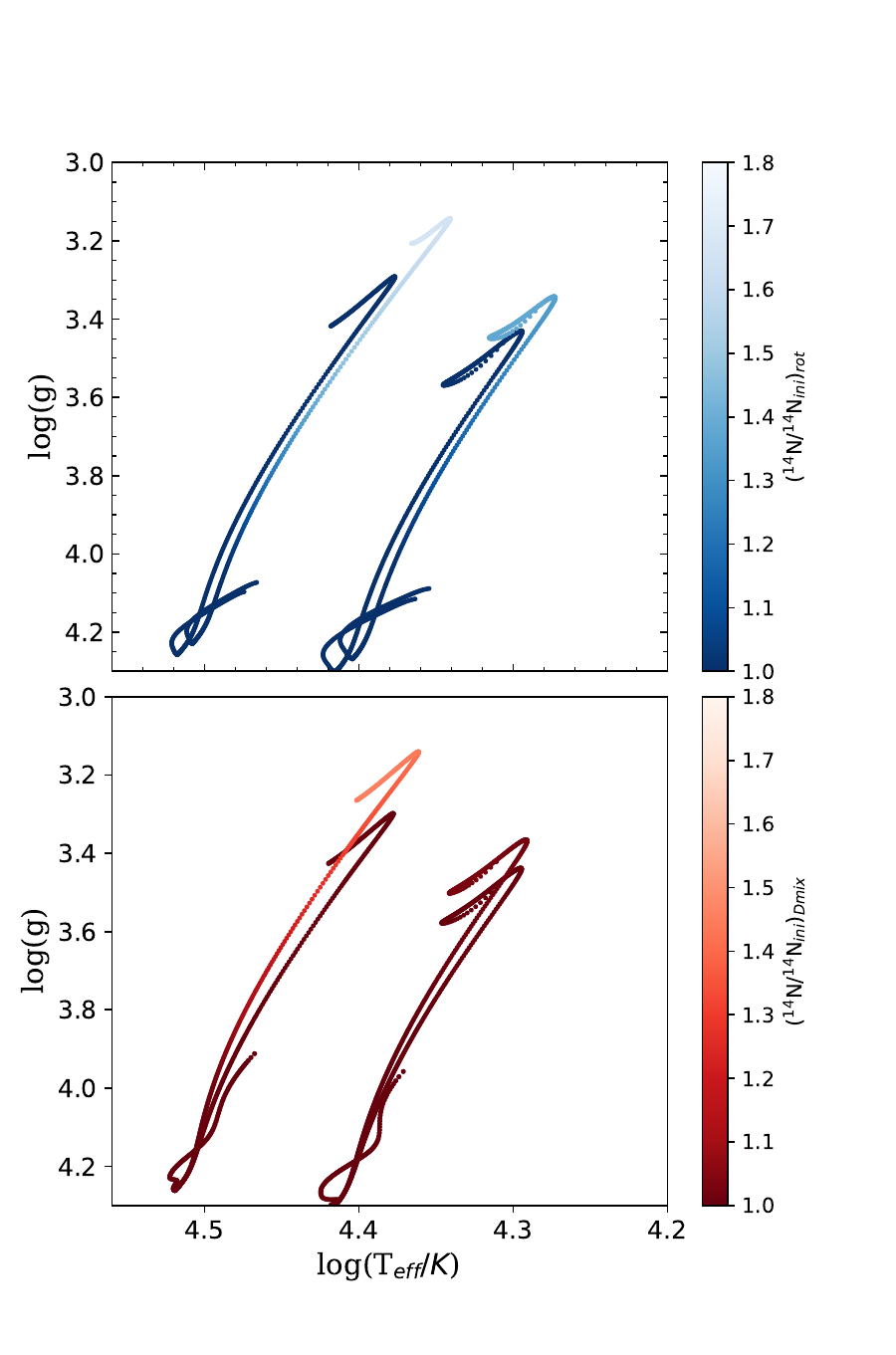}
     \caption{Tracks in the Kiel-diagram of 10 \msun{} (right) and 17 \msun{} (left) models, with the surface nitrogen compared to the initial surface abundance on the colour-scale, mind that a lighter colour indicates a larger amount of nitrogen on the surface. The top panel shows models with different levels of initial rotational velocity (blue scale), where left models of either set have an an initial rotational velocity of 15\% of the critical rotational velocity and the models more to the right have initial rotational velocity 40\% of the critical rotational velocity. These models do not include mixing based on IGWs. The bottom panel shows the models with different levels of envelope mixing based on IGWs and do not include the effects of rotational mixing. The tracks that branch off to lower log(g) are those with \Denv{}=5 for the 10\,\msun{} model and \Denv{}=6 for the 17\,\msun{} model. The other two tracks have \Denv{}=0.}
    \label{TheoreticallySpeaking}
\end{figure}

Most of the stars in the binary sample have a considerable surface rotational velocity (see column 5 of Table \ref{BinaryData}), which should also induce mixing in the stellar envelope according to the theory of rotational mixing \citep{Heger2000,Brott2011,Georgy2013}. In Fig.\,\ref{TheoreticallySpeaking} the change in the surface abundance of nitrogen is shown on the stellar evolution tracks in the Kiel diagram for the largest ($\sim$17\,\msun{}) and lowest ($\sim$10\,\msun{}) estimated masses of our binary sample\footnote{There are no abundances available for U Oph, thus we ignore the 5\,\msun{} models here.}. We compare the difference between enhanced envelope mixing induced by rotation but without IGW mixing (top panel) with the mixing due to IGWs but without rotational mixing (bottom panel), constructed as described in Section \ref{Method}. The models in the top panel have initial rotational velocities of 15$\%$ and 40$\%$ of the critical Keplerian rotational velocity, which corresponds to the lower and upper limits of the binary sample. The models with an initial rotational velocity of 15$\%$ the critical velocity do not show an increase in their nitrogen surface abundance, as the rotational mixing is not strong enough to have nitrogen reach the surface within the main-sequence lifetime. The models with an initial rotational velocity of 40$\%$ of the critical velocity do show an increase in the surface abundance of nitrogen (see also Table \ref{IGWvsRot}), with a factor of 1.7 for the 17\,\msun{} model and a factor of 1.4 for the 10\,\msun{} model, indicated by the lighter colour at the TAMS. 

The models with IGW envelope mixing have slightly lower levels of nitrogen surface enhancement. 
For the 10\,\msun{} model the surface enhancement of nitrogen is negligible, while for the 17\,\msun{} model the increase is a factor of 1.5 at the end of the main sequence.
From comparison of the upper and lower panel of the figure we find that our method for implementing the mixing based on IGWs from \citet{Michielsen2021} leads to slower nitrogen enhancement due to IGW mixing than the one due to rotational mixing from the theory by \citet{Heger2000} for the models rotating at 40$\%$ of the critical rate. 
Based on the models in Fig.\,\ref{TheoreticallySpeaking}, we expect an enhancement in the surface nitrogen abundance of both single stars and dEBs with a factor below two for 
\Denv{} < 6.

\begin{table}[]
    \centering
        \caption{The initial and final surface abundances of nitrogen for the models presented in Fig.\,\ref{TheoreticallySpeaking}}
    \begin{tabular}{ccccc}
         M$\rm_{ini}$&v$_{eq}$/v$_{crit}$& \Denv{}  &log$\epsilon$(N)$_{ini}$ &log$\epsilon$(N)$_{TAMS}$ \\
         \msun{} & $\%$ & - & - & -\\
         \hline
         10 &  15  &  0  & 7.79  &   7.79\\
         10 &  40  &  0  & 7.79  &   7.93\\
         10 &  0   &  0  & 7.79  &   7.79\\
         10 &  0   &  5  & 7.79  &   7.80\\
         17 &  15  &  0  & 7.79  &   7.79\\
         17 &  40  &  0  & 7.79  &   8.01\\
         17 &  0   &  0  & 7.79  &   7.79\\
         17 &  0   &  6  & 7.79  &   7.96\\
    \end{tabular}

    \label{IGWvsRot}
\end{table}

Unlike nitrogen, carbon and oxygen are not as sensitive to the internal mixing. For both of these elements the decrease in surface abundances is slower and less significant than the increase in nitrogen. Especially the change in oxygen is minimal because oxygen is depleted in the CNO-II cycle, which gets more active in more massive stars than we are modelling here. This means that independent of the profile of the internal mixing, for both the binary stars and the single pulsators, the level of internal mixing needed to change the $^{14}$N surface abundance significantly is rather high.

As a final note in this section, we point out that  \cite{Mombarg2025} recently investigated the impact of both rotational and IGW mixing on the surface abundance of nitrogen in OB-type stars covering the mass range of $[3,30]\,$M$_\odot$ and galactic to Small Magellanic Cloud metallicity. Their models are based on prescriptions of the joint effects of rotationally induced mixing (according to either the theory by \citet{ZahnChaboyer1992} or by \citet{Heger2000} as we used) and IGW mixing
from the simulations by \citet{Varghese2023}, without
taking into account the interaction between the waves and the rotation. The focus of their study was different than ours, as they did not consider observational constraints from  asteroseismology or dEBs. Rather, their aim was to calibrate IGW mixing from the most nitrogen enhanced single stars in the galaxy and Magellanic Clouds to deduce a calibration for the level of IGW envelope mixing. 
They find that IGW mixing levels derived from a 3rd degree polynomial scaling law of the stellar mass can reproduce the observed nitrogen enhancement at the surface of OB stars, while rotational mixing alone according to the measured surface rotation cannot. This conclusion is in agreement with the earlier findings of \citet{Aerts2014N14} focusing on a smaller mass range for galactic stars.  

For the rest of this paper, we focus on the differences in surface abundances between single B-type stars and dEBs with B-type components.

\subsection{Observational studies versus the CNO equilibrium}\label{CNOObservations}

\begin{figure}
     \includegraphics[width=\linewidth]{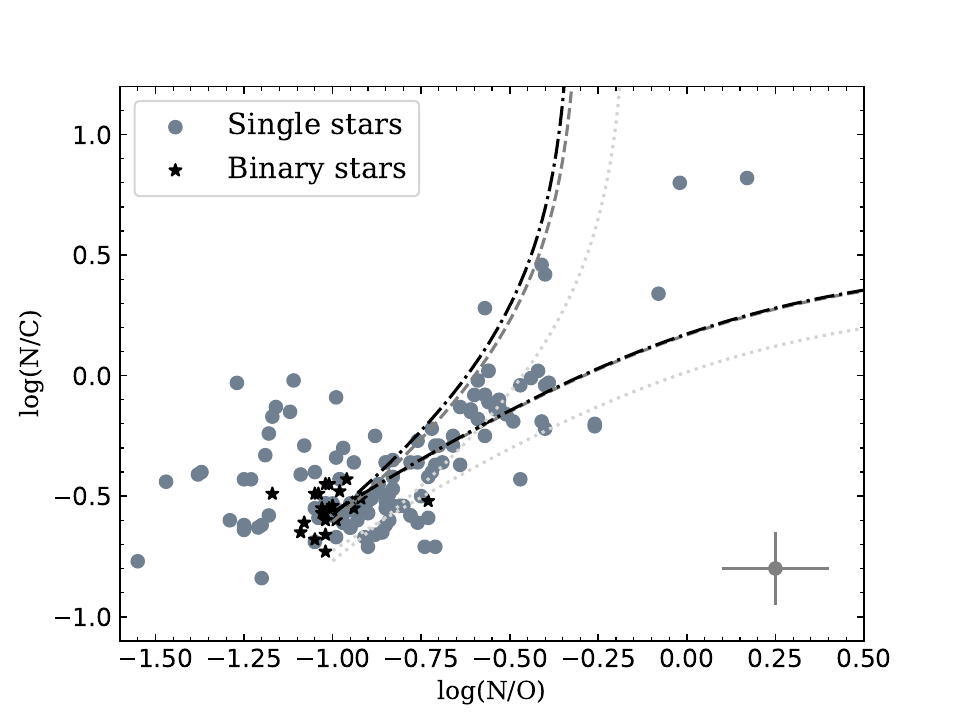}
    \caption{Observed values
     of N/C and N/O for single B-type stars from the literature (see text for references).
      The theoretical boundaries drawn by the lines are the same as in Fig.\,\ref{TheoreticallySpeaking}. The black stars indicate binary systems from \cite{Andrew2020}, \cite{Mayer2013}, \cite{Andrew2014B}, and \cite{Andrew2016}. The point in the lower right corner shows representative errors for all stars.}
    \label{CNOplotsSingle}
\end{figure}

To better understand the difference between the components of the binary systems and theoretical expectations, we check whether samples of presumed single B-type stars show the theoretically predicted behaviour in the CNO equilibrium. We collected several single star samples, mostly consisting of B-type stars, from the literature; (i) twenty slowly-rotating B-dwarfs from \cite{Morel2008}, (ii) 54 stars from \cite{Hunter2009} in the clusters NGC 6611, NGC 3293, and NGC 4755, (iii) eight B-type stars from the Orion star-forming region \citep{Nieva2011}, (vi) twenty B-type stars from the Solar neighbourhood from \cite{Nieva2012}, (iv) two early B-type stars from \cite{Fossati2015}, (v) 22 fast rotating massive stars from \cite{Cazorla2017}, (vi) twenty late O-type stars from \cite{Aschenbrenner2023}, and (vii) 55 Galactic early B-type stars from \cite{Jin2024} of which 15 overlap with the stars from NGC 3293 presented by \cite{Hunter2009}. 
Figure\,\ref{CNOplotsSingle} shows all the stars together in the CNO-equilibrium plot. Unlike the binary components (shown as the stars for reference), the single stars follow a diagonal as expected based on the theory underlying the CNO cycle. The scatter towards the right comes from the faster rotating stars from the samples of \cite{Cazorla2017} and \cite{Aschenbrenner2023}, while the scatter at the left side of the figure comes from the stars from \cite{Hunter2009}.

There are several reasons why the components of the binary systems and the single stars could behave differently: (i) there might be a systematic error in the effective temperature and the surface gravity of the stars which then in turn affects the abundances (and the mass discrepancy problem described earlier in Section \ref{discrepancy}), (ii) it can be due to a different method used for analysing the single and binary stars, which gives a different outcome for the surface abundances, (iii) it could be that tidal forces and the relatively fast enforced rotation affect the excitation of IGW at the core boundary and/or the angular momentum transport and thus influence the evolution of the internal rotation profile, which in turn may affect the excitation of IGW by the convective core as the stars evolve, following \citet{Varghese2024}. This would also imply that different levels of IGW mixing take place in the binaries compared to the single stars. 

\begin{figure*}
    \includegraphics[width=0.49\linewidth]{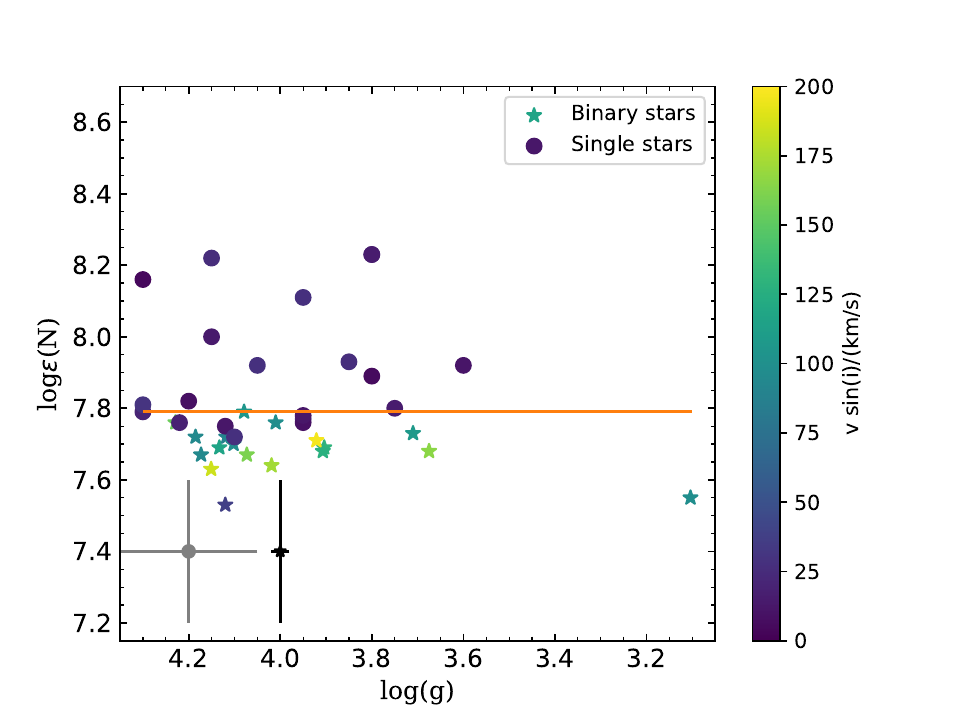}
    \includegraphics[width=0.49\linewidth]{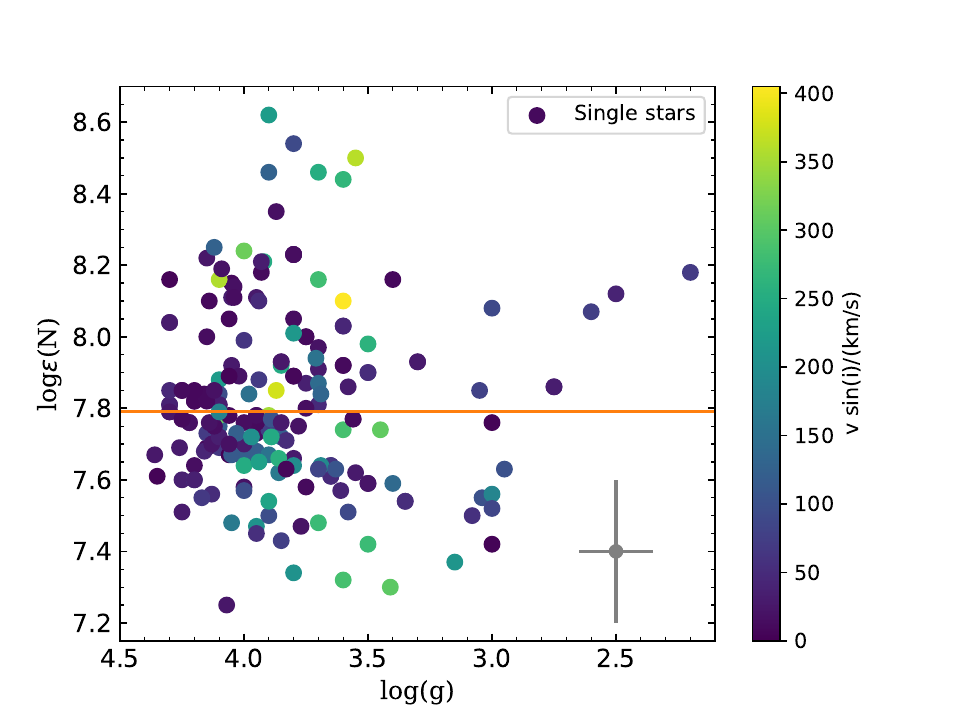}
    \caption{Left panel: surface abundance of N for the binaries (stars) and the single stars from the sample of \cite{Nieva2012} (circles) as a function of the surface gravity. The horizontal line is the average value of N from \cite{Nieva2012}, which we used to compute the stellar models in this work. The colour-scale shows the surface rotational velocity $v\sin\,i\ $ in km\,s$^{-1}$. Right panel: the same is shown for the various samples from the literature, with the addition of 33 stars from \cite{Aerts2014N14} and 28 stars from \cite{Morel2022}. Typical error bars for the single stars are shown in light grey in both panels and for the binaries in dark grey in the left panel.}
    \label{Surfaces}
\end{figure*}

A combination of all of the above reasons may cause the different behaviour between the single stars and dEBs.
 Since the changes in the CNO equilibrium mainly rely on changes in nitrogen we focus on that element from here on. In the left panel of Fig.\,\ref{Surfaces}, we compare the surface abundances of nitrogen from the binary systems with the single stars from the sample of \cite{Nieva2012}, since we used that elemental mixture for our models. For nitrogen, the scatter for the binaries is much smaller than for the single stars and the overall surface abundance  of the dEBs is lower. While this was expected based on Figs\,\ref{CNOLimits} and \ref{CNOplotsSingle}, there is 
 no clear trend of the nitrogen surface abundance with the surface gravity, as would be expected based on Fig.\,\ref{TheoreticallySpeaking}. Also, there are a few stars with high surface abundances of nitrogen and a high $\log (g)$, but with a low rotational velocity. While this might be due to the star being the product of a merger \citep[see, e.g.,][]{Schneider2020}, it points to a combined occurrence of slow rotation and nitrogen enrichment, in line with \citet{Varghese2024}.

 For some of the samples, all stars are slow rotators and yet the nitrogen abundance is scattered. For the samples that include fast rotating stars, there is no clear trend between the nitrogen surface abundance and the rotational velocity which was already found by \cite{Aerts2014N14}. Based on stellar evolution theory, one would expect the stars with lower values of the surface gravity to have a higher surface abundance since mixing takes time to occur in the radiative envelopes of these stars. As the sample is small to evaluate this expectation, we show the same quantities in 
the right panel but for all single star samples without the binaries. A few extra stars from the previously mentioned samples were added, as they lacked either C or O, but have an estimation for N. We also added 33 stars from \cite{Aerts2014N14} and 28 stars from \cite{Morel2022}. The sample of \cite{Morel2022} has an overlap with both \cite{Hunter2009} and \cite{Jin2024}, and we removed the overlapping stars from the latter two studies, leaving 31 stars from \cite{Hunter2009} and 48 stars from \cite{Jin2024}. The studies from \cite{Aerts2014N14} and \cite{Morel2022} were not included in Fig.\,\ref{CNOplotsSingle} because the former only presents nitrogen abundances and the latter does not include oxygen. Figure\,\ref{Surfaces} does not reveal a clear correlation between the surface abundance of nitrogen and the surface gravity. The Pearson correlation coefficient of the data is less than 1 $\%$. There is also no clear connection between the rotational velocity and the nitrogen abundance at the surface, but we do observe that the few most evolved stars 
with $\log\,g < 2.9$ are all slow rotators and enriched in nitrogen.

As a last remark of this section, we recall that the surface abundances of the stellar evolution models do not only depend on the internal mixing of the models. They also depend on the initial chemical mixture of elements which are put into the model. Along with the metallicity, the chosen mixture has an impact on the evolution of the individual abundances, the speed with which they change, and the final values at the end of the main sequence. The chemical mixture determines how a star appears as enriched, as it determines the starting point of the chemical evolution within the star \citep[cf.\ ][]{Mombarg2025}. From an observational point of view, it matters where the star is located in the Galaxy, as the metallicity and the chemical mixture change depending on the location within the disk.

\section{Uncertainties from spectroscopic analyses}\label{Observations}

While the binary components considered in this work are found at different evolutionary stages along the main sequence and display a diversity in rotation rates, the surface nitrogen abundances reported in the literature are consistent with cosmic standard values and show no significant dispersion (see Fig.\,~\ref{CNOLimits}). This observation is at odds with theoretical predictions based on classical rotational mixing, which foresees surface nitrogen enrichment in stars that have evolved past the mid-main-sequence stage (see Figure~\ref{TheoreticallySpeaking}).

In this section, we investigate the potential role of observational uncertainties and data processing in the apparent lack of response in the measured surface nitrogen abundance to the combined effects of rotation (or any other source of mixing in the envelope) and evolution in the binary sample. To that end, we focus on the most extreme object in the sample — the primary component of the V380 Cyg system. This star is reported to be located near or just beyond the TAMS for its mass and has an equatorial rotational velocity of approximately 100 km\,s$^{-1}$. According to predictions from stellar evolution models, such a star should exhibit surface signatures of the CNO cycle, particularly in the form of nitrogen enrichment. Contrary to expectations, however, the observed nitrogen abundance is consistent with the cosmic standard value reported by \citet{Nieva2012} and is indistinguishable from the surface nitrogen abundance found in much younger stars of similar mass.

The more evolved primary component of V380 Cyg has been reported in the literature to exhibit a large microturbulent velocity field, with a microturbulence parameter of $\xi = 15 \pm 1$ km\,s$^{-1}$ \citep[see, e.g.,][]{Andrew2014}. \citet{Andrew2020} demonstrated that an inconsistent treatment of microturbulence in the spectroscopic analysis of binaries, where the effect is included in the line formation code but neglected in the atmospheric model calculations, can lead to an overestimation of the effective temperature. This occurs because: (i) a large microturbulent velocity field in the stellar atmosphere increases the importance of the turbulent pressure; (ii) to conserve total pressure in the stellar atmosphere, the contribution of the gas pressure term must decrease, resulting in weaker and narrower hydrogen lines than under normal conditions (i.e., when microturbulence is weak); (iii) under the assumption of a fixed $\log\,g$, which is typical for dEBs where this parameter is computed from the star’s mass and radius, the effect of weaker and narrower hydrogen lines is most easily matched by increasing the effective temperature in model fits. In the case of the primary component of V380 Cyg, \citet{Andrew2020} reported this effect to be as strong as 8\%, meaning that the inferred effective temperature could be overestimated by up to 8\% relative to its actual value in the atmosphere of the star.

An incorrectly inferred effective temperature will misplace a star on the Kiel diagram and is expected to impact the derived surface chemical composition. For example, the V380 Cyg system is well known in the context of the mass discrepancy problem. Depending on the study, a difference of up to 40\% has been reported between the dynamical and evolutionary masses for the primary component \citep{Andrew2014}. When plotted on the $\log(T_{\rm eff})$–$\log(g)$ Kiel diagram, this mass discrepancy manifests itself in a star appearing hotter (and, to some extent, more evolved, though this depends on the amount of near-core mixing assumed in single-star evolutionary models) than predicted by the evolutionary track corresponding to its dynamical mass. Thus, if the effective temperature is inaccurately determined, it may contribute to both the mass discrepancy and the apparent lack of variation in surface nitrogen abundance observed in dEBs.

\subsection{Pressure balance and atmospheric parameters}\label{sec2}
Here, we test the hypothesis that neglecting turbulent pressure in the calculation of stellar atmosphere models leads to an overestimation of the star’s effective temperature, $T_{\rm eff}$. To quantify this effect, we simulate the emergent spectrum of a B-type star with the following parameters: $T_{\rm eff} = 20,000$~K, $\log g = 3.1$~dex, [M/H] = 0.0~dex (solar), $v\,\sin\,i = 100$~km\,s$^{-1}$, and a microturbulent velocity parameter $\xi = 15$~km\,s$^{-1}$. The latter value is set both in the model atmosphere (i.e., accounting for the contribution of turbulent pressure) and in the line formation code. To enhance realism, we add Poisson noise to the synthetic spectrum to mimic a signal-to-noise ratio (S/N) of 250.

\begin{figure*}
    \centering
    \includegraphics[width=\linewidth]{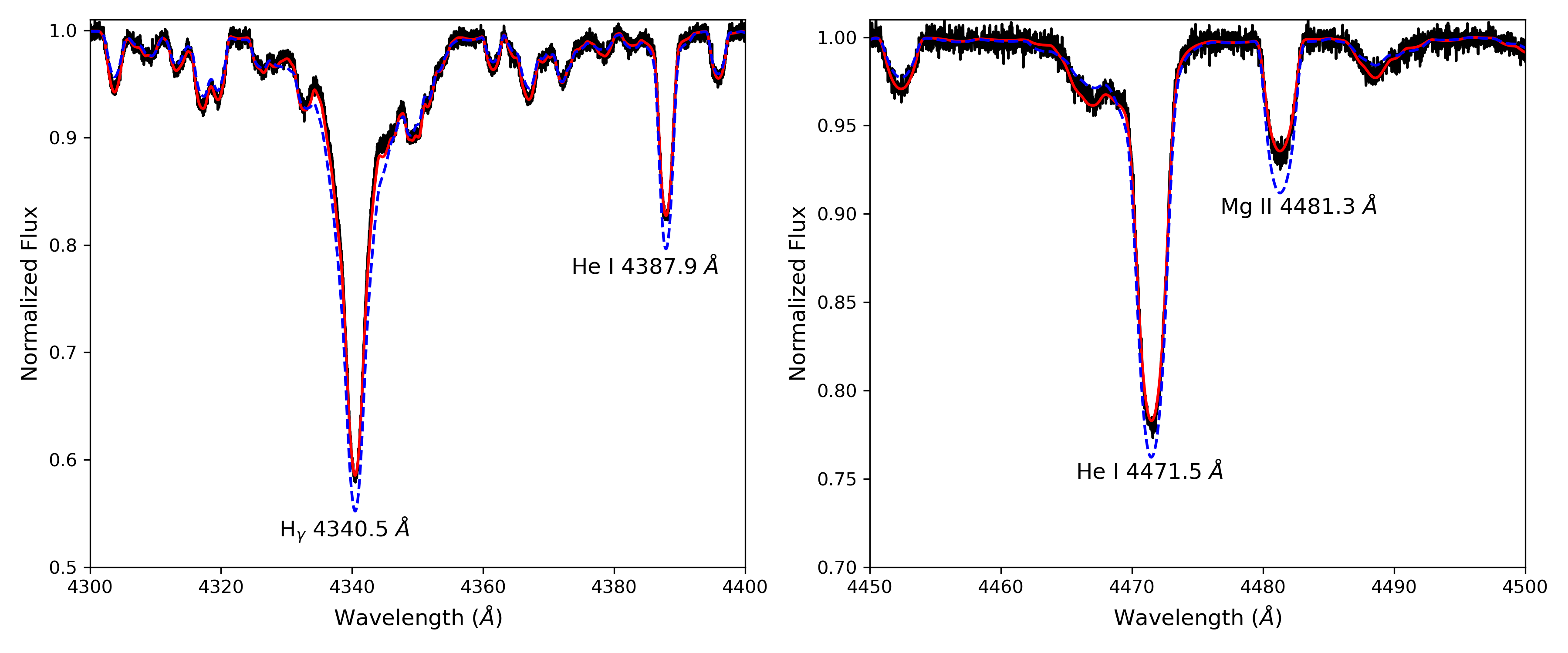}
    \caption{Comparison between the artificial observed spectrum of the primary component of V380 Cyg (black solid line) and the best-fit synthetic model (red solid line). For reference, a model spectrum with the same parameters as the artificial observed spectrum — except for $\xi = 2$~km\,s$^{-1}$ in the model atmosphere — is shown as a blue dashed line. See text for details.}
    \label{fig:PressureTeff}
\end{figure*}

This artificial spectrum is then analysed with the GSSP software package \citep{Andrew2015} using a standard grid of stellar atmosphere models computed with a microturbulent velocity of $\xi = 2$~km\,s$^{-1}$. This effectively neglects the influence of turbulent pressure on the atmospheric structure. During the analysis, we fix the surface gravity $\log\,(g)$ to 3.1~dex, thereby replicating the case of the primary component of V380 Cyg, where this parameter is derived from the star’s dynamical mass and radius.

Figure~\ref{fig:PressureTeff} shows the quality of the fit between the artificial observed spectrum (black solid line; hereafter “observations”) and the best-fitting model (red solid line), which yields the following parameters: $T_{\rm eff} = 21,590 \pm 150$~K, [M/H] = $-0.03 \pm 0.05$~dex, $v\,\sin\,i = 98.5 \pm 3.5$~km\,s$^{-1}$, and $\xi = 14.4 \pm 0.8$~km\,s$^{-1}$. For comparison, we also show a model corresponding to the input parameters of the artificial spectrum, but computed from a standard atmosphere model with $\xi = 2$~km\,s$^{-1}$ (blue dashed line). This model fails to reproduce the observations; in particular, the H$_\gamma$ hydrogen line appears too broad and overly strong, as do the lines of neutral helium and singly ionised magnesium. In contrast, increasing $T_{\rm eff}$ from 20,000~K to approximately 21,600~K (an 8\% increase) enables an accurate reproduction of hydrogen and helium lines, as well as most metallic lines (compare the red and black lines in Fig.\,~\ref{fig:PressureTeff}). This result confirms the findings of \citet{Andrew2020}, namely that neglecting the turbulent (or any additional) pressure term in the atmosphere model leads to an overestimation of the effective temperature by approximately 8\% in the temperature regime of B-type stars.

\begin{figure*}
    \centering
    \includegraphics[scale=0.51]{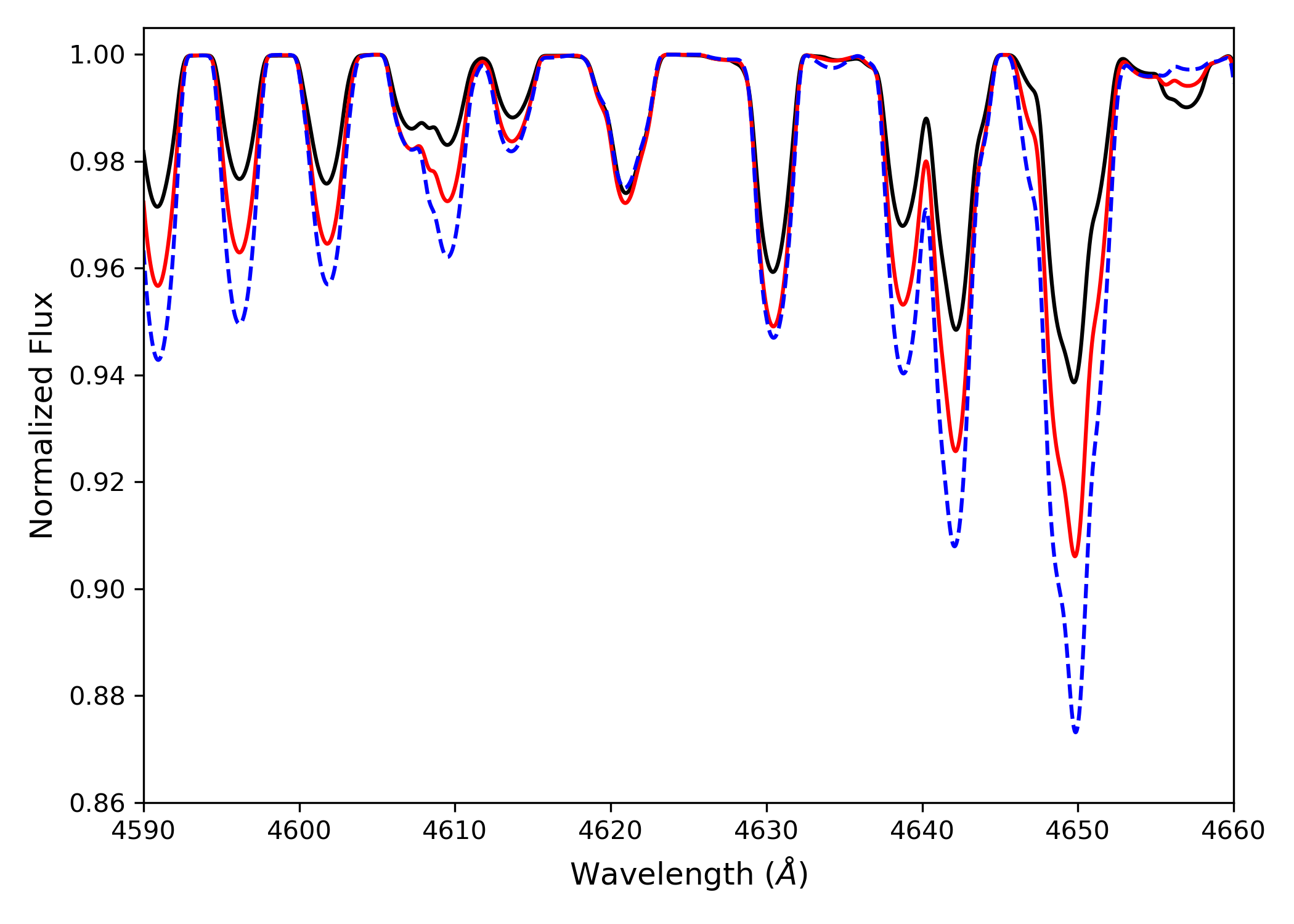}
    \includegraphics[scale=0.51]{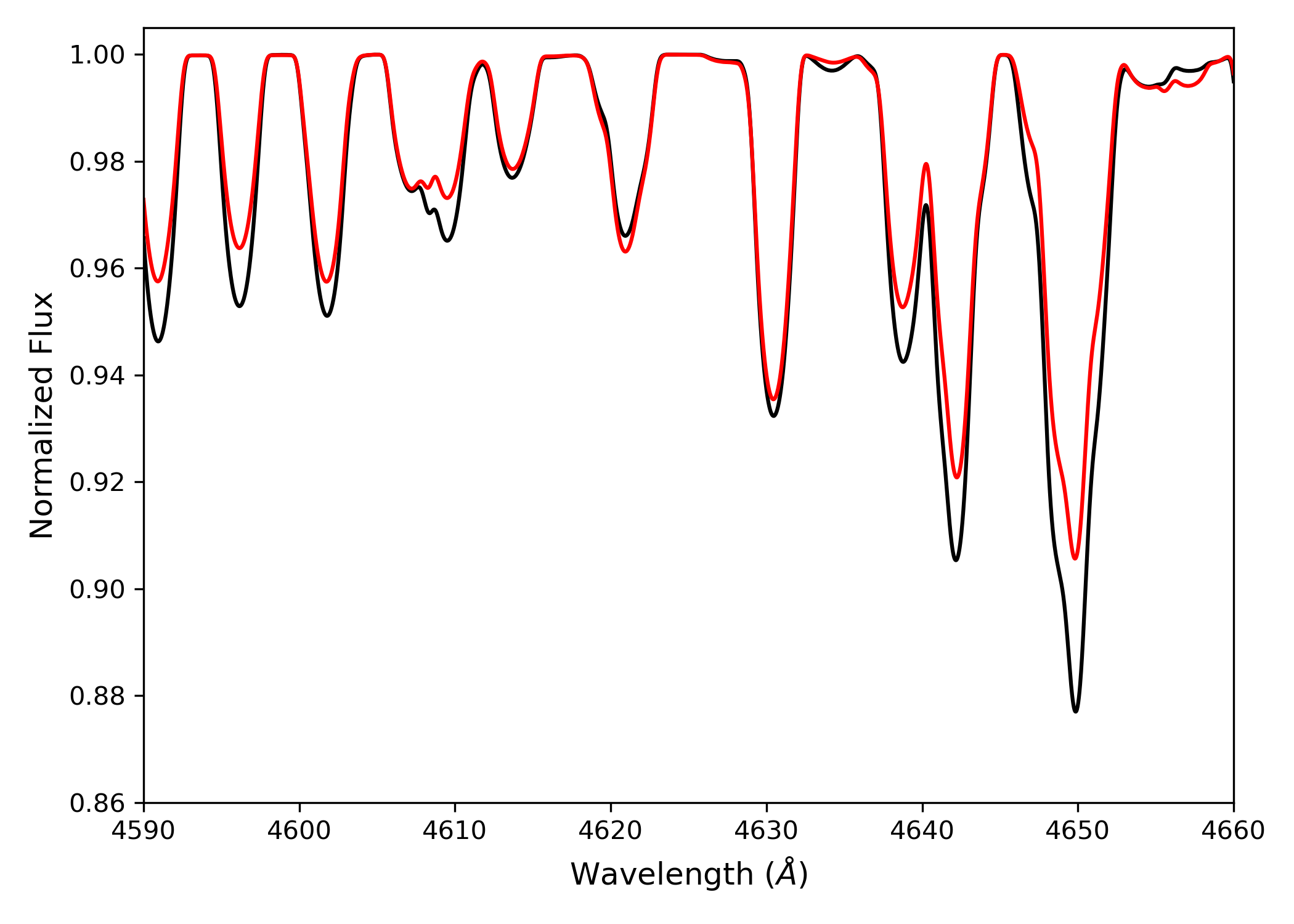}
    \caption{Effect of variations in atmospheric parameters on nitrogen lines in the atmosphere of a V380 Cyg-like star. Left: Black, red, and blue lines represent effective temperatures of 18\,500 K, 20\,000 K, and 21\,500 K, respectively. Right: Black and red lines show the effect of changing the microturbulent velocity in the atmospheric model from 15 km~s$^{-1}$ to 2 km~s$^{-1}$, respectively.}
    \label{fig:NitrogenLines}
\end{figure*}

\subsection{Nitrogen abundance from the artificial spectrum}
We now proceed with a test designed to reveal the impact of an inaccurately determined effective temperature on the inferred surface nitrogen abundance. The left panel of Fig.\,~\ref{fig:NitrogenLines} shows a wavelength range rich in singly ionised nitrogen lines and illustrates their response to variations in effective temperature by approximately 8\% in either direction from the baseline value of 20\,000~K. The depths of the N II lines increase with increasing effective temperature, with the relative increase being somewhat more pronounced at lower temperatures.

Since neglecting the turbulent pressure in stellar atmosphere models is the root cause of the incorrectly inferred effective temperature, it is important to consider the effect of the increased atmospheric microturbulent velocity on the behaviour of nitrogen lines. The right panel of Fig.\,~\ref{fig:NitrogenLines} shows the effect of changing the microturbulence in the atmospheric model from 2~km\,s$^{-1}$ (red line) to 15~km\,s$^{-1}$ (black line). The  nitrogen lines systematically increase in depth as the turbulent pressure in the stellar atmosphere increases.
Moreover, a comparison between the left and right panels of Fig.\,~\ref{fig:NitrogenLines} reveals that the magnitude of the effect of increasing the effective temperature from 20\,000~K to 21\,500~K (red versus blue line in the left panel) is comparable to that of decreasing the microturbulent velocity from 15~km\,s$^{-1}$ to 2~km\,s$^{-1}$ (black versus red line in the right panel).

In practice, for the specific case of the primary component of V380 Cyg, we expect the effects of an incorrectly inferred effective temperature and the neglect of turbulent pressure in the atmospheric model to cancel each other with respect to the inference of the surface nitrogen abundance. To test this hypothesis quantitatively, we proceed with the analysis of the artificial spectrum of the primary component of V380 Cyg, this time focusing on the determination of the surface nitrogen abundance. In this analysis, we fix the atmospheric parameters according to the solution derived in Section~\ref{sec2}, namely: $T_{\rm eff} = 21\,590 \pm 150$~K, [M/H] = $-0.03 \pm 0.05$~dex, $v \sin i = 98.5 \pm 3.5$~km\,s$^{-1}$, and $\xi = 14.4 \pm 0.8$~km\,s$^{-1}$. We focus on several wavelength regions rich in nitrogen spectral lines: 3990–4060~\AA, 4152–4255~\AA, 4410–4455~\AA, 4590–4700~\AA, 4770–4815~\AA, 4980–5051~\AA, 5660–5720~\AA, and 5920–5960~\AA.

We remind the reader that the artificial observed spectrum was generated using the following set of parameters: $T_{\rm eff} = 20\,000$~K, $\log g = 3.1$~dex, [M/H] = 0.0~dex (solar), $v\sin i = 100$~km\,s$^{-1}$, and a microturbulent velocity parameter $\xi = 15$~km\,s$^{-1}$. Our analysis is based on a grid of atmospheric models computed with the standard microturbulence value of $\xi = 2$~km\,s$^{-1}$. This approach is intended to simulate the combined effect of an incorrectly inferred effective temperature and the neglect of turbulent pressure in the atmosphere model on the derived surface nitrogen abundance. The analysis yields a good quality of fit, with a reduced $\chi^2$ value close to unity across all considered wavelength intervals, and a nitrogen abundance of $-4.26 \pm 0.10$~dex. This value is consistent with the overall metallicity assumed in the artificial spectrum, supporting our earlier hypothesis that the opposing effects of increased effective temperature and neglected turbulent pressure effectively cancel out in the determination of the nitrogen abundance. 

Our conclusion of this section is therefore that the different behaviour between the single 
 B stars and the dEB components is not due to biases in the observational analyses, as illustrated by our test for V380\,Cyg, which is the most extreme case among our dEB sample.

\section{Summary and conclusions}\label{Conclusions}

The main goal of this work was to determine whether the observed trends of the surface nitrogen abundance in components of dEBs can be explained by wave-induced mixing in the envelope rather than by rotation-induced mixing as is commonly assumed. The answer to our research question is affirmative. To come to this conclusion, we developed a line of research summarized as follows.

We first determined the amount of envelope mixing caused by IGW needed to match theoretical models of B-type stars with observations of such stars. Our models are based on mixing induced by IGW excited stochastically by core convection \citep{Rogers2013}, leading to a mixing profile according to the particle diffusion simulations by \citet{Rogers2017} for a 3\,M$_\odot$ ZAMS star and by \citet{Varghese2023,Varghese2024} for a range of masses and evolutionary stages representing main-sequence B stars.
We concluded that relatively large amounts of IGW mixing at the level of \Denv{}=5-6 near the convective core, adopting the mixing profile in Fig.\,\ref{Mixing}, are required. 
Such high levels are required to reach the observed effective temperatures and luminosities for the higher mass components of the binary systems in our sample.
This conclusion is in agreement with \cite{Andrew2020}, though these authors only considered  additional mixing in the transition layer between the convective core and the radiative envelope as indicated in Fig.\,\ref{Mixing}. Since 
the main subject of their investigation was the mass discrepancy problem, they focused on near-core mixing
affecting the convective core mass during the evolution of their target stars. Our focus was on the surface abundances requiring envelope mixing to be active in order to explain measurements of surface nitrogen measured in B stars.
The amount of envelope mixing required to transport CNO processed nitrogen to the surface in our models is in line with the maximum amount found in previous asteroseismic studies of single B-type pulsators \citep[see, e.g.,][]{Pedersen2021}. 
We point out that our essential conclusions would not change if we had taken into account the variation in the IGW mixing profiles for a variety of masses or evolutionary stages found by \citet{Varghese2023}, because we took the efficiency level at the stitching point as a free parameter. \citet{Varghese2023} find the profiles resulting from their simulations to be rather similar in shape except close to the TAMS, but most of our sample stars are still far away from the TAMS.  The only nuance would be that (avoiding) surface enrichment would require slightly (lower) higher levels of 
D$_{\rm env}$.

To confront our model predictions with observations, we questioned whether such realistic levels of envelope mixing are physically justifiable in the context of dEBs, which offer model-independent stellar parameters and have not yet undergone binary interactions. For such objects, comparisons with theoretical predictions of single-star models are meaningful.
We thus assembled surface abundance measurements of B-type star components in dEBs in the literature, to check whether they show any enhancement of nitrogen and accompanying changes in carbon and oxygen. The assembled sample includes stars located from near the ZAMS to close to the TAMS, with the binary V380\,Cyg as prototypical example. Since this binary hosts a near-TAMS primary, envelope mixing had sufficient time to result in nitrogen enhancement if it were efficient (see Figure~\ref{TheoreticallySpeaking}). V380\,Cyg 
has an equatorial rotation velocity close to 100~km\,s$^{-1}$ and has been shown to exhibit stochastic low-frequency variability, presumably due to IGWs \citep{Andrew2014}. Indeed,  \cite{Nadya2024} have shown that its observed low-frequency variability cannot be due to a subsurface convective zone given its mass. V380\,Cyg does not have enhanced surface nitrogen. Hence IGW mixing, if present in its envelope, cannot be efficient
and must be at a level \Denv{} < 6. 

We further highlighted that none of the B stars in the assembled sample of dEBs, regardless of surface rotation rate or evolutionary stage, show an enhancement in their surface nitrogen. Rather, they all  cluster in the N/C versus N/O diagram in the position where envelope mixing is not efficient throughout their evolution (see Fig.\,\ref{CNOLimits}). This behaviour was already noticed before by \citet{Pavlovski2018,Pavlovski2023}.
From our new theoretical models quantified by \Denv{}, we conclude that, if envelope mixing is caused by IGW, all the B stars in the dEB sample must have \Denv{} < 6 in order to explain the observed lack of enhancement. This is a main conclusion of our study, which we showed to be unaffected by uncertainties in the observational spectroscopic analyses that led to the measurement of the surface nitrogen abundances of the dEBs.

To improve our understanding of envelope mixing, we subsequently also considered various samples of single B-type stars from the literature. While these samples, containing both slow and fast rotating stars, do show the expected trend in the CNO abundances caused by envelope mixing (see Fig.\,\ref{CNOplotsSingle}), their surface abundance of nitrogen does not show a correlation with either the surface gravity or the projected surface rotation velocity (see Fig.\,\ref{Surfaces}). This finding is consistent with the earlier results in \citet{Aerts2014N14}, who found no correlation between the surface nitrogen abundance and  equatorial rotation velocity for a sample of 33 single B-type stars with both these measurements available. It led to the conclusion that rotational mixing alone, according to its standard theory, cannot explain the observed nitrogen enhancements.

\citet{Varghese2023} found that IGW mixing in non-rotating single B-type stars covering a mass range of [3,20]\,M$_\odot$ is most efficient in the earliest evolutionary phases. Moreover, the more massive stars have more effective IGW mixing. Considering rotating models up to 70\% of the critical rate,  \citet{Varghese2024} subsequently showed the efficiency of IGW mixing to decrease strongly with increasing near-core rotation rate. Most of the stars in our sample of dEBs have fast rotation throughout their evolution enforced by the tidal forces due to their binarity. Their IGW mixing can therefore not be effective according to \citet{Varghese2024}'s simulations, because of the enforced rotation. On the other hand, the sample of single stars reveals the occurrence of nitrogen enhancement in the few most evolved stars, which had sufficient time to transport the CNO processed material to the surface, and in the slow rotators for which IGW mixing is effective throughout the evolution. 
We add two more facts to these findings: i) 
the measured surface nitrogen abundances have quite large uncertainties, and ii)  asteroseismic modelling showed that main-sequence B-type stars with measurements of differential rotation typically 
rotate below 50\% of their critical rate  \citep{Aerts2025}
and have core-to-envelope rotation ratios below 4 \citep{Dupret2004,Briquet2007,Suarez2009,Triana2015,Burssens2023,Fritzewski2025,Vanlaer2025}. Hence the efficiency of the excitation of IGW at their core boundary can be meaningfully assessed from their surface rotation.

Putting all the above aspects together we conclude that evolutionary models with IGW mixing profiles based on the simulations by \citet{Varghese2023,Varghese2024} can potentially explain the surface nitrogen properties of both single and binary B-type stars, depending on their value for \Denv{}.
Rotation does play a role, but not in a matter predicted by the standard theory of rotational mixing. Rather, rotation decreases or prevents the efficiency in the
excitation of IGWs at the core boundary, while the excited IGWs are responsible for the efficiency of the wave-induced envelope mixing that may act together with rotational mixing during the main-sequence evolution \citep{Mombarg2025}.

 Asteroseismology of pulsating dEBs is ideally suited to scrutinize our hypothesis further. For the time being, the samples of such objects with asteroseismic modelling are small \citep{Aerts2024}. Moreover, we need such modelling to be done for stars having also proper measurements of surface abundances. The sample of 14,000 new eclipsing binaries with OBAF-type components from \cite{IJspeert2024}, several of which having pulsating components, may constitute an important resource for the next steps in this research.

\begin{acknowledgements}
The authors thank the MESA team for making their modules publicly available. We also thank Kresimir Pavlovski for valuable discussions regarding the surface abundances of binary stars. The research leading to these results has received funding from the Research Foundation Flanders (FWO) under grant agreement G089422N, as well as from the BELgian federal Science Policy Office (BELSPO) through the PRODEX grant for PLATO. 
CA further acknowledges financial support from the European Research Council (ERC) under the Horizon Europe programme (Synergy Grant agreement N°101071505: 4D-STAR). While partially funded by the European Union, views and opinions expressed are however those of the author(s) only and do not necessarily reflect those of the European Union or the European Research Council. Neither the European Union nor the granting authority can be held responsible for them.

\end{acknowledgements}
\bibliographystyle{aa} 
\bibliography{references} 
\end{document}